\documentclass[aps,showpacs,superscriptaddress,reprint,amsmath,amssymb]{revtex4-1}
\usepackage{graphicx,color}
\usepackage{dcolumn}
\usepackage{bm}

\begin{document}

\title{Defect-induced magnetism in SiC probed by nuclear magnetic resonance}

\author{Z. T. Zhang}
\email[Corresponding author. E-mail: ]{z.zhang@hzdr.de}
\affiliation{Institute of Ion Beam Physics and Materials Research, Helmholtz-Zentrum Dresden-Rossendorf, D-01314 Dresden, Germany}
\affiliation{Hochfeld-Magnetlabor Dresden (HLD-EMFL), Helmholtz-Zentrum Dresden-Rossendorf, D-01314 Dresden, Germany}

\author{D. Dmytriieva}
\author{S. Molatta}
\author{J. Wosnitza}
\affiliation{Hochfeld-Magnetlabor Dresden (HLD-EMFL), Helmholtz-Zentrum Dresden-Rossendorf, D-01314 Dresden, Germany}
\affiliation{TU Dresden, D-01062 Dresden, Germany}

\author{Yutian Wang}
\affiliation{Institute of Ion Beam Physics and Materials Research, Helmholtz-Zentrum Dresden-Rossendorf, D-01314 Dresden, Germany}
\affiliation{School of Microelectronics, Xidian University and Key Laboratory of Wide Band-Gap Semiconductor Materials and Devices, Xi\textquoteright an 710071, China}

\author{M. Helm}
\affiliation{Institute of Ion Beam Physics and Materials Research, Helmholtz-Zentrum Dresden-Rossendorf, D-01314 Dresden, Germany}
\affiliation{TU Dresden, D-01062 Dresden, Germany}

\author{Shengqiang Zhou}
\affiliation{Institute of Ion Beam Physics and Materials Research, Helmholtz-Zentrum Dresden-Rossendorf, D-01314 Dresden, Germany}

\author{H. K\"{u}hne}
\email[Corresponding author. E-mail: ]{h.kuehne@hzdr.de}
\affiliation{Hochfeld-Magnetlabor Dresden (HLD-EMFL), Helmholtz-Zentrum Dresden-Rossendorf, D-01314 Dresden, Germany}

\date{\today}

\begin{abstract}
	
We give evidence for intrinsic, defect-induced bulk paramagnetism in SiC by means of $^{13}$C and $^{29}$Si nuclear magnetic resonance (NMR) spectroscopy. The temperature dependence of the internal dipole-field distribution, probed by the spin part of the NMR Knight shift and the spectral linewidth, follows the Curie law and scales very well with the macroscopic DC susceptibility. In order to quantitatively analyze the NMR spectra, a microscopic model based on dipole-dipole interactions was developed. The very good agreement between these simulations and the NMR data establishes a direct relation between the frequency distribution of the spectral intensity and the corresponding real-space volumes of nuclear spins. The presented approach by NMR can be applied to a variety of similar materials and, thus, opens a new avenue for the microscopic exploration and exploitation of diluted bulk magnetism in semiconductors. 

\end{abstract}
\maketitle

Defect-induced magnetism is a fascinating topic that is generating strong research interest, not only since it promotes progress in practical applications, such as in quantum bits and spintronics, but also because it raises fundamental questions about the basic understanding of the magnetism in a material without any partially filled $3d$ or $4f$ shells \cite{SiC-qubit1,SiC-qubit2,SiC-neutron-YuLiu,SiC-Ne-LiZhou,SiC-proton-shandong,SiC-Ne-WangZhou, SiC-Ne-WangZhou-jap,SiC-Xe-WangZhou-srep,SiC-neutron-WangZhou,graphite-pi-electron,graphite-C-imp, graphite-RTFM-ptdefect,graphite-H-He,NC-nano-diamond, C60-film,HfO2,TiO2,ZnO,MoS2,MoS2-2}. Such magnetism was observed in a wide range of materials, e.g., in SiC \cite{SiC-neutron-YuLiu,SiC-Ne-LiZhou,SiC-proton-shandong,SiC-Ne-WangZhou,SiC-Ne-WangZhou-jap, SiC-Xe-WangZhou-srep,SiC-neutron-WangZhou}, carbon-based materials \cite{graphite-pi-electron,graphite-C-imp,graphite-RTFM-ptdefect,graphite-H-He,NC-nano-diamond,C60-film}, oxides \cite{HfO2,TiO2,ZnO}, and MoS$_{2}$ \cite{MoS2,MoS2-2}. Different schemes have been proposed to uncover its origin. For example, first-principles calculations revealed that the local moments in neutron-irradiated SiC arise from the $sp$ states of divacancy ($V_{\mathrm{Si}}V_{\mathrm{C}}$) defects \cite{SiC-neutron-YuLiu}, whereas Ohldag \emph{et al.} showed that the magnetic order in proton irradiated graphite is due to the carbon $\pi$-electron system \cite{graphite-pi-electron}. Clearly, this fundamental question is still far from being settled.
 
Experimentally, the weak magnetic signal, often just slightly above the detection limit of a SQUID magnetometer \cite{GaNGd-weak-signal,TiO2,SiC-neutron-YuLiu,SiC-Ne-LiZhou}, requires much care to preclude extrinsic factors, such as magnetic impurities or contaminations \cite{purity-substrate1,purity-substrate2,purity-substrate3,purity-substrate4,purity-substrate5}. As a local-probe technique, nuclear magnetic resonance (NMR) provides a way to get insight into the intrinsic magnetic properties of a material. In fact, an NMR spectrum directly maps the internal magnetic-field distribution, sampled at the atomic positions of the addressed nuclear moments. For example, NMR was used to determine the  magnetic hyperfine field in ferromagnetic graphite \cite{graphite-NMR-srep}. Recently, we showed that ferromagnetism in neutron-irradiated SiC (NI-SiC) exists in an intermediate fluence range \cite{SiC-neutron-WangZhou}. However, paramagnetism always occurs with the amplitude scaling with the irradiation fluence.

In this work, we use $^{13}$C and $^{29}$Si NMR spectroscopy as well as numerical simulations to study the defect-induced paramagnetism in neutron irradiated 6\emph{H}-SiC with a relatively high defect concentration. Consistent with macroscopic DC susceptibility measurements, the temperature dependence of the NMR shift can well be described by the Curie law for both nuclear isotopes, in full compatibility with intrinsic paramagnetism. The Curie behavior is also followed by the linewidth of the NMR spectra, indicating a growing width of the internal dipole-field distribution stemming from the local moments upon cooling. In order to achieve a quantitative understanding of the experimental NMR spectra, we developed a microscopic model based on dipole-dipole interactions. Very good agreement between the experiments and simulations is achieved, and an anisotropic distribution of the local moments is inferred. Further, we show that the intrinsic paramagnetism in NI-SiC can well be described in a local-moment picture. 

\begin{figure}[tbp]
	\centering
	\includegraphics[scale=1]{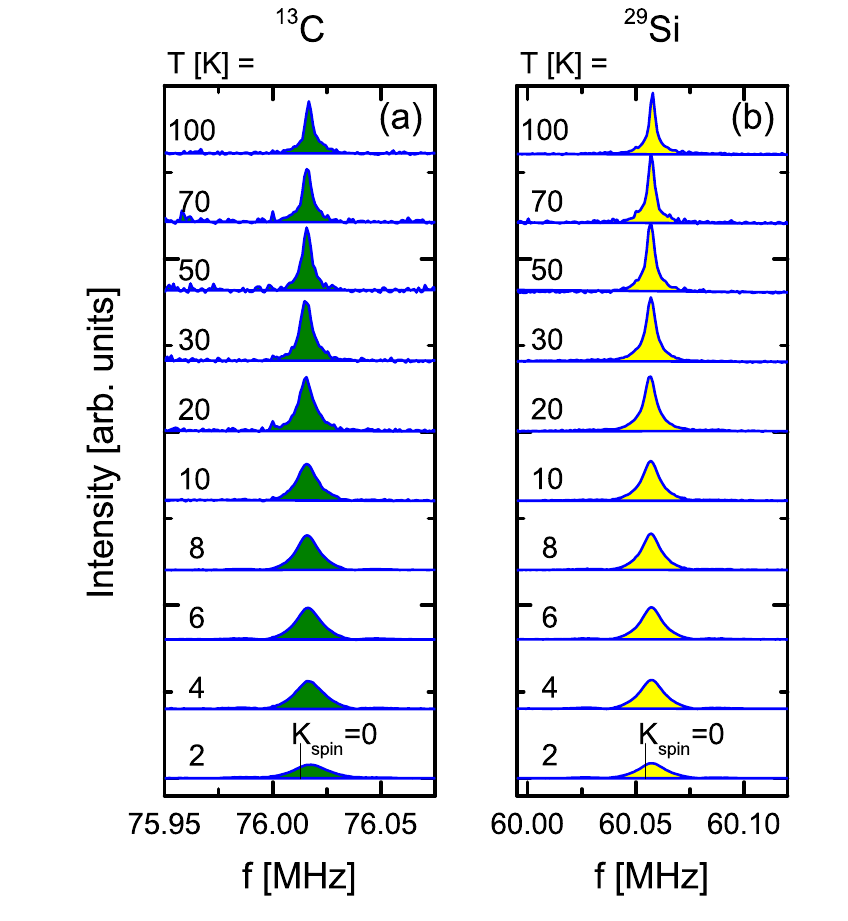}
	\caption{\label{Fig_Spectra} Experimental (a) $^{13}$C and (b) $^{29}$Si NMR spectra for NI-SiC at different temperatures. The vertical markers over the spectra at 2 K label the frequency where $K_{spin}=0$.}
\end{figure}

Commercial semi-insulating 6\emph{H}-SiC (001) single-crystal wafers were irradiated with neutrons at the reactor BER II (Position DBVK) at Helmholtz-Zentrum Berlin \cite{Neutron-exp-detial1}. During irradiation, the temperature of the samples was less than 50 $^{\circ}$C \cite{Neutron-exp-detial2}. The paramagnetic sample used for our NMR measurements was irradiated for 150 hours with the fluence reaching $3.12\times10^{19}$ cm$^{-2}$ (only fast neutrons). The $^{29}$Si and $^{13}$C NMR spectra were acquired with a Hahn spin-echo pulse sequence at temperatures between 2 and 100 K at a magnetic field of $\mu_{0}H=7.100$ T. For the temperature-dependent measurements, the magnetic field was applied in the wafer plane. Angle-dependent measurements were performed at $T=2$ and 100 K. The macroscopic magnetization was measured using a SQUID-VSM magnetometer (Quantum Design). All of the experiments were performed on the same NI-SiC sample.

\begin{figure}[tbp]
	\centering
	\includegraphics[scale=0.9]{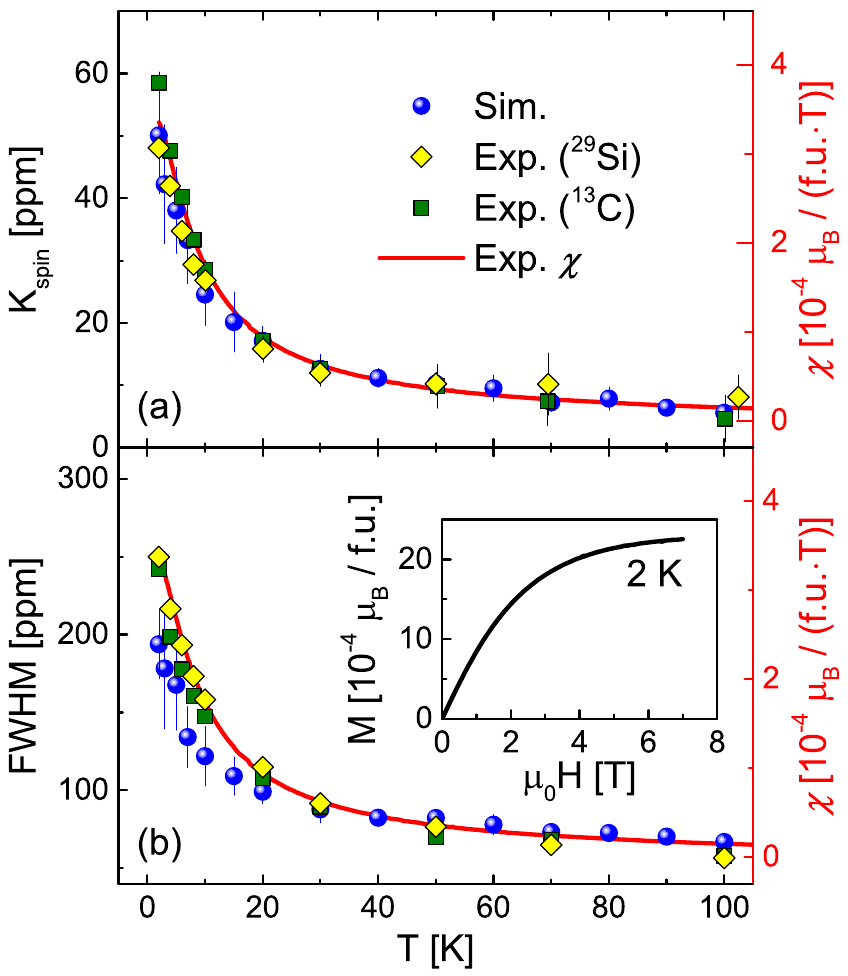}
	\caption{\label{Fig_Kspin} Temperature dependence of (a) the shift $K_{spin}$ and (b) the linewidth of the $^{13}$C and $^{29}$Si NMR spectra, both in comparison to the simulated values for 
	$N_c=3$ and $\beta=20$. Details of the simulation are given in the text. The red curves in (a) and (b) represent the macroscopic DC susceptibility $\chi(T)$ measured at 7 T. The inset of (b) shows the magnetization as a function of applied field at 2 K.}
\end{figure}


Figure \ref{Fig_Spectra} shows the experimental $^{13}$C and $^{29}$Si NMR spectra at temperatures between 2 and 100 K. Since both $^{13}$C and $^{29}$Si have a nuclear angular momentum of $I=1/2$, only one spectral line from the transition $I_{z}=-1/2$ to $+1/2$ is observed. The full width at half maximum (FWHM) of the spectra at 100 K is less than 5 kHz, indicating the very high quality of the single-crystalline sample. As the sample is cooled, the FWHM becomes larger, indicating the growing width of the magnetic dipole-field distribution stemming from the localized defect moments.

The NMR shift $K$ is defined as $K=(f_{res}-f_{0})/f_{0}$, where $f_{0}={\gamma_{n}}\mu_{0}H/2\pi$ is the Larmor frequency of a bare nucleus with a gyromagnetic ratio ${\gamma_{n}}$ in a magnetic field $\mu_{0}H$, and $f_{res}$ is the NMR frequency. In the present case, the NMR shift $K$, calculated as the first moment of the experimental NMR spectra, is the combination of a spin part and a temperature-independent orbital contribution: $K=K_{spin}+K_{orb}$. The spin part $K_{spin}=A_{hf}\chi_{spin}$ is proportional to the uniform susceptibility of the electronic spins and to the hyperfine coupling constant, $A_{hf}$, between the nuclear and electronic spin moments.

In the present work, $K_{orb}$ (the NMR shift where $K_{spin}=0$) is obtained from the very sharp $^{13}$C and $^{29}$Si NMR spectra of a pristine SiC sample at 100 K. Using the $^{63}$Cu NMR signal of the sample coil as an in-situ reference, we determined $K_{orb}(^{13}C) = 114 \pm 5$ ppm and $K_{orb}(^{29}Si) = 117 \pm 5$ ppm, respectively. Thereby, it was possible to extract $K_{spin}$ for both isotopes [Fig. \ref{Fig_Kspin}(a)]. The temperature dependence of $K_{spin}$ follows the Curie law, in very good agreement with the macroscopic susceptibility $\chi(T)$. The paramagnetism is also indicated by the Curie behavior of the FWHM for both isotopes [Fig. \ref{Fig_Kspin}(b)], which quantitatively reflects the temperature-dependent width of the dipole-field distribution from the local moments. These results give clear evidence of intrinsic paramagnetism in the NI-SiC sample, and are consistent with our previous findings \cite{SiC-neutron-WangZhou}.
 
For an estimate of the average hyperfine coupling constant, the defect concentration is needed. Noting that the magnetic moments are almost fully polarized at $T=2$ K and $\mu_{0}H=7$ T [inset of Fig. \ref{Fig_Kspin}(b)], we evaluate the average defect concentration to be $\approx 0.00117$ f.u.$^{-1}$, using the total measured magnetization and 2 $\mu_{\mathrm{B}}$ per defect as input parameters. In fact, it was reported that the defect moments introduced by neutron irradiation are mostly divacancies ($V_{\mathrm{Si}}-V_{\mathrm{C}}$), with an $S=1$ state and a moment of 2 $\mu_{\mathrm{B}}$, as was shown by magnetometry and ESR experiments \cite{SiC-neutron-WangZhou,SiC-neutron-YuLiu}. The scaling factor between $K_{spin}$ and $\chi$ gives an estimate of the average hyperfine coupling constant $A_{hf}$ $\approx$ 0.13--0.16 f.u.$\times$T/$\mu_{\mathrm{B}}$, or  $\approx$ 1.5--1.9$\times10^{-4}$ T/$\mu_{\mathrm{B}}$, respectively. Such small values are in agreement with dipole-dipole interactions over an average distance of a few nanometers.

\begin{figure}[tbp]
	\centering
	\includegraphics[scale=1]{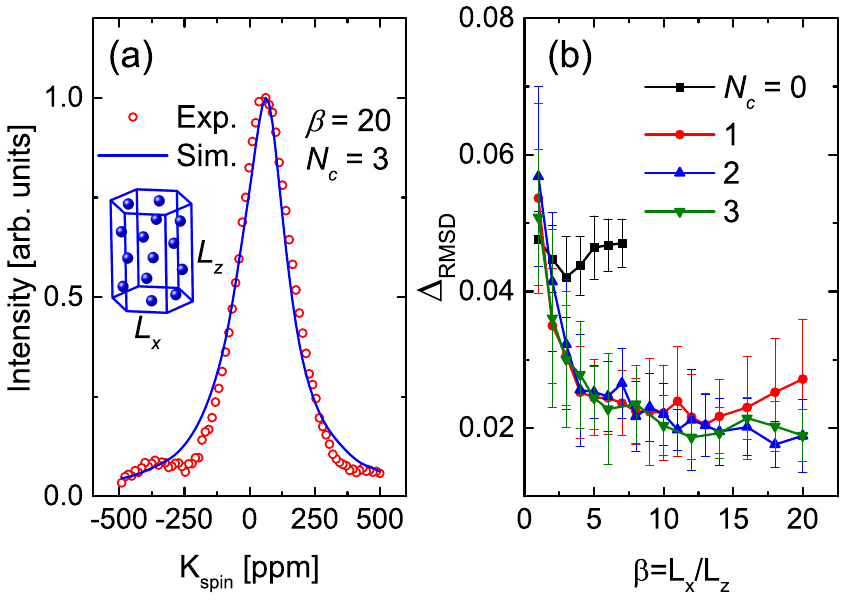}
	\caption{\label{Fig_sim_spec} (a) Experimental $^{13}$C NMR spectrum measured at 2 K with field applied in plane (red open circles) and simulated NMR spectrum for $N_c=3$ and $\beta=20$ (blue solid line). (b) Difference $\Delta_{\mathrm{RMSD}}$ between the experimental spectrum at 2 K and the simulation as a function of the compression factor $\beta$.}
\end{figure}

In order to obtain a quantitative understanding of the experimental NMR spectra, we developed a microscopic model based on the real-space distribution of magnetic dipole fields, stemming from the randomly positioned defect moments.
The simulated NMR spectra are obtained by sampling the magnetic hyperfine fields at the individual nuclear sites over a real-space volume $V_{0}$.
A schematic representation of $V_{0}$ with hexagonal symmetry is shown in the inset of Fig. \ref{Fig_sim_spec}(a). For each site, the NMR spectrum is constructed using the Lorentzian line-shape function,
\begin{equation}
	f(x)=1/\left[4(\frac{x-K^{i}}{\Delta})^{2}+1\right],
\end{equation}
where $K^{i}$ is the calculated NMR shift for site $i$ and $\Delta$ is the intrinsic spectral linewidth. The shift $K^{i}$ is calculated from the summation of the dipole fields that stem from all surrounding localized moments in the considered volume,
\begin{equation}
	\mathbf{B}_{dip.}^{i}=\frac{\mu_{0}}{4\pi}\sum_{j}^{\mathrm{surr.}}\mathbf{m}\cdot
	\left(\frac{3\mathbf{r}_{ij}\mathbf{r}_{ij}}{r_{ij}^{5}}-\frac{1}{r_{ij}^{3}}\right),
\end{equation}
where $\mathbf{m}$ is the magnetic moment of a defect and the summation range of $j$, defining the size of the magnetic moment structure, is limited by a cut-off parameter $N_{c}$. E.g., for $N_{c}=0$, only the defect moments in $V_{0}$ are taken into account; for  $N_{c}=1$ also the nearest-neighbour volumes with size identical to $V_{0}$ are considered.

In addition to the cut-off parameter $N_{c}$, a parameter $\beta=L_{x}/L_{z}$, i.e. a compression factor of $V_{0}$ in $z$-direction ($c$ axis), is introduced, since, in the present case of pure dipole-dipole interactions, a non-zero shift of the first spectral moment can only result from an anisotropy of the moment distribution. Figure \ref{Fig_sim_spec}(a) displays the simulated and experimental $^{13}$C spectra at 2 K, and Fig. \ref{Fig_sim_spec}(b) shows the $\beta$-dependent $\Delta_{\mathrm{RMSD}}$ (the root-mean-square deviation between the simulated and experimental spectra).

\begin{figure}[tbp]
	\centering
	\includegraphics[scale=1]{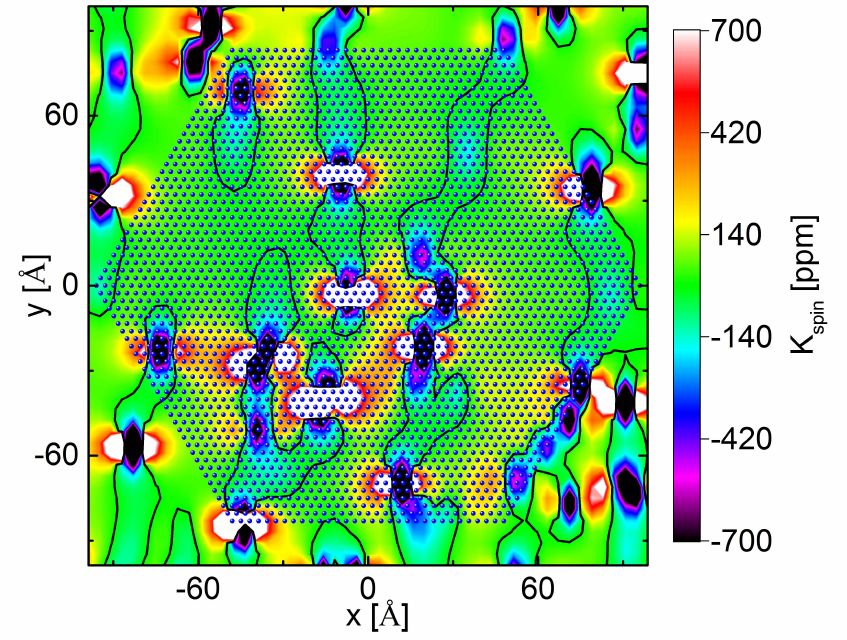}
	\caption{\label{Fig_SpatialK} Contour plot of the simulated distribution of $K_{spin}$ as function of the real-space coordinates $(x,y,z=0)$ for $N_c=3$ and $\beta=20$, with the magnetic field applied along the $x$ direction. The randomly positioned divacancy moments appear as extended objects with high color constrast, blue points label the position of the carbon sites, black solid lines label the positions where $K_{spin}=0.$} 
\end{figure}

Very good agreement between the simulated and experimental NMR spectra is achieved when $\Delta_{\mathrm{RMSD}}$ is minimized with respect to $N
_{c}$ and $\beta$.
Since the dipole fields decay only over long distances, no good agreement can be reached for the case $N_{c}=0$, which takes only the local moments inside $V_{0}$ into account. When $N_{c}$ is increased up to 1 or larger, a convergence is approached. $\Delta_{\mathrm{RMSD}}$ decreases rapidly for $\beta<5$. For $5<\beta<15$, $\Delta_{\mathrm{RMSD}}$ decreases much weaker, and then levels off for higher $\beta$. The required large value of $\beta$ indicates an anisotropic distribution of the defects, i.e., a shorter average distance between the moments along the $z$ direction. 
The underlying mechanism for the anisotropic defect distribution is likely
related to the symmetry of the crystal structure and the resulting anisotropy of the atomic displacement energies \cite{Devanathan2000}. A quantitative statement could be obtained by detailed simulations of the displacement dynamics and is subject to future studies.

\begin{figure}[tbp]
	\centering
	\includegraphics[scale=1]{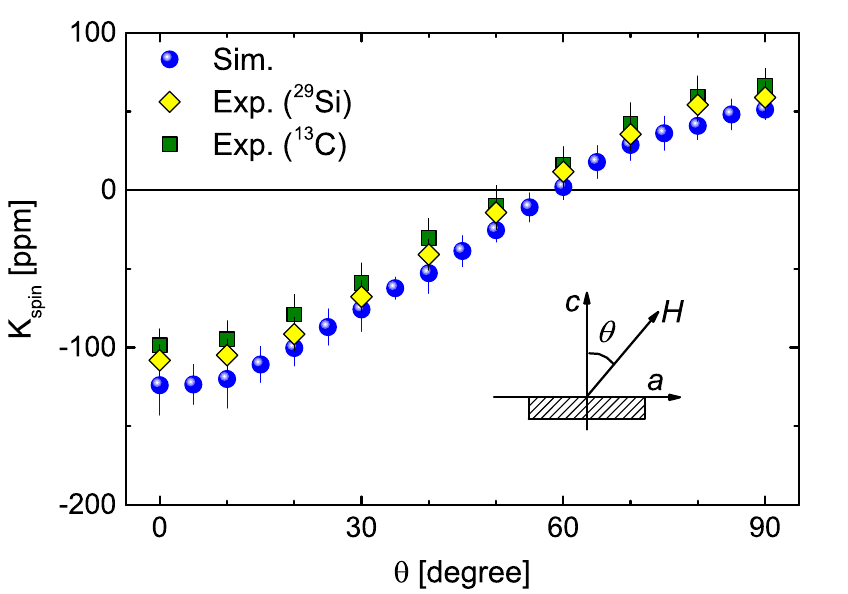}
	\caption{\label{Fig_sim_angleK} Angular dependent spin part of the $^{29}$Si (yellow diamonds) and $^{13}$C (green squares) NMR shifts, $K_{spin} = K_{2\mathrm{K}}-K_{100\mathrm{K}}$, and simulated NMR shift for $N_c=3$ and $\beta=20$ (blue circles). The inset depicts the definition of the angle $\theta$.}
\end{figure}

In Fig. \ref{Fig_SpatialK}, a  contour plot of a typical simulated real-space distribution of $K_{spin}$ is shown. The NMR spectra are identical to a sampling of this distribution with a finite step size, given by the underlying sublattice of either the  $^{13}$C or $^{29}$Si nuclear moments as local-probe magnetometers. In consequence,
we obtain basically the same results from $^{13}$C or $^{29}$Si NMR, differing only by the values of the respective nuclear gyromagnetic ratios as well as the local orbital contributions. 
The major volume fraction (green color), contributing dominantly to the spectral weight, yields values of $K_{spin}$ between about $\pm140$ ppm. The local moments are indicated by the red-white-black singular points, where $K_{spin}$ is large but contributes little to the spectral weight due to the very low volume fraction. 
The whole volume is distributed into positive and negative regions of $K_{spin}$, the black contour lines indicate $K_{spin}=0$. With the given value of $\beta = 20$, 
positive values of $K_{spin}$ yield the dominant volume fraction, resulting in a positive first-moment shift of the simulated spectrum. Moreover, a majority of the nuclear sites has an average distance of a few nanometers from the nearest local moment. This supports the statement that the average hyperfine coupling agrees with the dipole-dipole interactions over a few nanometers. We note that our simulations
do not consider any clustering or cooperative mechanisms of the defect moments, such as  ferromagnetism, which 
was observed in small volume fractions of 
4H SiC after irradiation with a low fluence level of neutrons \cite{SiC-neutron-WangZhou}. For such phenomena, there is either no indication by our experimental results or the corresponding, small volume fractions are expected to give only a negligible contribution to the NMR spectra. 

Having established a very good agreement between the experimental NMR spectra and the microscopic simulation at 2 K, we now discuss the comparison of the simulation to the temperature-dependent experimental results. At each temperature point, a spectrum was simulated and $K_{spin}$ as well as the linewidth were extracted. For this, all simulation parameters were fixed to the values obtained from the comparison at 2 K. The only temperature-dependent parameter is amplitude of the defect moment $\mathbf{m}$, which we take to follow the Curie behavior, in compatibility with the DC susceptibility. The results 
of this simulation are compared to the experimental values in Fig. \ref{Fig_Kspin}.
For $K_{spin}$, the temperature dependence of the experimental data is in excellent
agreement with the simulations. For the spectral linewidth, 
again, we find an excellent agreement at high temperatures, and a small residual disagreement at low temperatures.

To further confirm and explore the anisotropy introduced by a finite value of $\beta$, we considered the angular dependence of the NMR shift with respect to the applied external field, which defines the polarization axis of the defect moments.
Our simulations reveal an angular dependence of the NMR shift, presented in Fig. \ref{Fig_sim_angleK}. Depending on the orientation of the applied magnetic field, the shift of the first spectral moment can be positive or negative. To confirm these results experimentally, we performed angular dependent NMR measurements at 2 and 100 K. Since $K_{orb}$ is anisotropic and temperature independent, we take $K_{2\mathrm{K}}-K_{100\mathrm{K}}$ as a good approximation of $K_{spin,2\mathrm{K}}$, noting that $K_{spin,2\mathrm{K}}$ is $\approx 50$ times larger than $K_{spin,100\mathrm{K}}$. The measured angular-dependent $^{29}$Si and $^{13}$C NMR shifts are shown in Fig. \ref{Fig_sim_angleK}. 
Again, the simulations show very good agreement with the experimental values. This confirms the anisotropic distribution of the local moments, inferred from the analysis of the temperature-dependent NMR shift data.

In summary, we used a combined approach by NMR spectroscopy and numerical simulations to investigate the defect-induced magnetism in NI-SiC. The intrinsic nature of bulk paramagnetism is revealed by the Curie behavior of the temperature-dependent NMR frequency shift as well as the spectroscopic linewidth. A microscopic simulation of the real-space dipole-field distribution, generated by the defect moments, was developed. The very good agreement with the experimental data establishes a direct relation between the frequency distribution of the spectral intensity and the corresponding real-space volumes of nuclear spins. Perspectively, this allows for a controlled, volume-selective manipulation of nuclear spins by narrow-band excitations within the NMR spectrum.
The presented approach by NMR spectroscopy and microscopic simulations can be used for a broad range of similar material compounds and, thus, opens a new avenue for the exploration of dilute magnetism in semiconductors and applications of quantum bits and spintronics.

\begin{acknowledgments}
The neutron irradiation was done at the Helmholtz-Zentrum Berlin f\"{u}r Materialien und Energie by Gregor Bukalis. The project is supported by the Helmholtz Association (VH-PD-146). Z.T.Z. was financially supported by the National Natural Science Foundation of China (Grant No. 11304321) and by the International Postdoctoral Exchange Fellowship Program 2013 (Grant No. 20130025). Further, support by the HLD at HZDR, a member of the European Magnetic Field Laboratory, and by the Deutsche Forschungsgemeinschaft (DFG) through the Research Training Group GRK 1621 is gratefully acknowledged.
\end{acknowledgments}

\end{document}